# Valence Band Dispersion Measurements of Perovskite Single Crystal with Angle-resolved Photoemission Spectroscopy

Congcong Wang, Benjamin R. Ecker, Yongli Gao[*]

Department of Physics and Astronomy, University of Rochester, Rochester, NY 14627, USA

Haotong Wei, Jinsong Huang

Department of Mechanical and Materials Engineering, University of Nebraska-Lincoln, Lincoln, Nebraska 68588, USA.

Jian-Qiao Meng

School of Physics and Electronics, Central South University, Changsha, Hunan 410083, PR China.

* ygao@pas.rochester.edu

Abstract

The electronic structure of the cleaved perovskite ($CH_3NH_3PbBr_3$) single crystal was studied in an ultra-high vacuum (UHV) system by angle-resolved photoemission spectroscopy (ARPES) and inverse photoelectron spectroscopy (IPES). Highly reproducible dispersive features of the valence bands were observed with nice symmetry about the Brillouin zone center and boundaries. The largest dispersion width was found to be ~0.73 eV and ~0.98 eV along the $\Gamma X$ and $\Gamma M$ directions, respectively. The effective mass of the holes was estimated to be ~0.59 $m_0$. The quality of the surface was verified by atomic force microscopy (AFM) and scanning electron microscope (SEM). The elemental composition was investigated by high resolution x-ray photoelectron spectroscopy (XPS). The experimental electronic structure shows a good agreement with the theoretical calculation.

Introduction

Hybrid organic-inorganic halide perovskites have first attracted intense attention for their potential applications in high-performance photovoltaic devices.[1-5] After only a few years of active research, the power conversion efficiency of the devices have reached 22.1%[6]. More importantly, the halide perovskites' uses have expanded beyond just photovoltaic devices to a variety of other equally important applications such as photo detectors[5], light emitting diodes (LED)[7] and lasers[8]. While the material growth processes[9-12], electronic density of states[13-15] and the stability[16-18] have been widely studied, the material's underlying intrinsic carrier transport property remain poorly understood. They have so far been limited to either theoretical modelling[19] or indirectly acquisition[20] from transistors made from polycrystalline film where real intrinsic property could be masked. Indeed for every application listed above, such characteristic is the key to a fundamental understanding and prediction of the actual device performance.

Studies of band structure are very important for understanding the electronic properties and the transport characteristics. Theoretical calculations and combined experimental studies have been performed to investigate the band structure of perovskite[21-28], yet few direct measurements of band dispersion, effective mass and hole mobility determining the transport characteristics have been made. Angle-resolved photoemission spectroscopy (ARPES) is an ideal technique to directly measure the band structure. To focus on the intrinsic carrier dynamics, freshly cleaved single crystals have been used to reduce any possible extrinsic factors, such as impurities and grain boundaries.

In this manuscript, we report our investigation on the band dispersion of high quality $CH_3NH_3PbBr_3$ single crystal with ARPES. The band widths are measured to be 0.73 and 0.98 eV along $\Gamma X$ and $\Gamma M$, respectively. Together with inverse photoelectron spectroscopy (IPES), the band gap was measured to be 2.3 eV. The slight *n*-type crystal was cleaved in ultra-high vacuum (UHV) system and the composition was investigated with high resolution x-ray photoelectron spectroscopy (XPS). The quality of the surface was investigated by atomic force microscopy (AFM) and scanning electron microscope (SEM).

Experimental

Sample Preparation

The high quality methylammonium lead bromide perovskite single crystals were synthesized by solution-processed anti-solvent growth method as described in ref. 5. Briefly, 0.64 M $PbBr_2$ and 0.8 M methylamine bromine were dissolved into 5 ml *N,N*-Dimethylformamide solution in a small vial. Then the vial was sealed with foil, but leaving a small hole to let dichloromethane slowly get in. Dichloromethane was employed as anti-solvent to precipitate the single crystals. Finally, the vial was stored in the atmosphere of dichloromethane, and $CH_3NH_3PbBr_3$ single crystal slowly grew into big in two days. Each experiment was performed on a freshly cleaved sample at room temperature. Especially for ARPES, IPES and XPS measurements, the crystals were cleaved in an UHV system (*in situ*). For AFM, SEM and energy dispersive spectroscopy (EDS) measurements, the samples were cleaved *ex-situ*. The average size of the crystals is ~6 mm $\times$ 6 mm $\times$ 3.5 mm.

Characterizations

The morphology and the initial elemental analysis of the as-cleaved sample were observed by an NTMDT AFM Microscope and a Zeiss Auriga SEM with a built-in EDS system. Powder XRD measurements were performed with a Rigaku D/Max-B X-ray diffractometer with Bragg–Brentano parafocusing geometry, a diffracted beam monochromator, and a conventional cobalt target X-ray tube set to 40 kV and 30 mA.

Photoelectron Spectroscopy Measurements

The ARPES measurements were performed on perovskite single crystal (001) surface using a VG ESCA Lab UHV system equipped with a He I (21.2 eV) gas discharge lamp. The base pressure of the spectrometer chamber is typically $8 \times 10^{-11}$ Torr. The typical instrumental energy resolution for ultraviolet photoelectron spectroscopy (UPS) measurements ranges from ~0.03 to 0.2 eV with photon energy dispersion of less than 20 meV. The UV light spot size on the sample is about 1 mm in diameter. The detector angular resolution is ~0.5°. The surface composition information was measured by high resolution XPS with a monochromatic Al $K\alpha$ source (1486.6 eV) in an UHV chamber with a base pressure of $1 \times 10^{-10}$ Torr. The energy resolution of XPS is about 0.6 eV. The x-ray spot size on the sample is about 0.1 mm in diameter. The IPES spectra were recorded using a custom-made spectrometer, composed of a commercial Kimball Physics ELG-2 electron gun and a band pass photon detector. The photon detector worked in the isochromatic mode centered at a fixed energy of 9.8 eV. The combined resolution (electron and photon) of the IPES spectrometer was determined to be ~0.6 eV.

Results and Discussion

Fig. 1a shows the as-grown perovskite single crystal (001) surface with an average area of 6 mm × 6 mm. The thickness of the sample is ~3.5 mm. Powder XRD of the grounded crystal confirmed the cubic crystal structure of the sample (Fig. S1). The reciprocal lattice of the crystal unit cell is shown in Fig. 1b, where $\Gamma X$, $\Gamma Y$, and $\Gamma M$ directions corresponds to the *a*-axis, *b*-axis (Fig. 1a) and the face diagonal direction, respectively. Due to the crystal symmetry, $\Gamma X$, $\Gamma Y$ and $\Gamma Z$ directions are all degenerate. Therefore, later ARPES measurements were only performed along the $\Gamma X$ and $\Gamma M$ directions. The three-dimensional (3D) and two-dimensional (2D) AFM images were taken at three different locations (Fig. 1c, 1d, S2) to illustrate the high quality of the surface, which is a prerequisite for achieving accurate band structure results from ARPES. The average root mean square (RMS) roughness of the three positions is as low as 4.273 nm, indicating a very smooth surface. The surface quality was also imaged by SEM, and the built-in EDS system confirmed the existence of Pb and Br (Fig. S3).

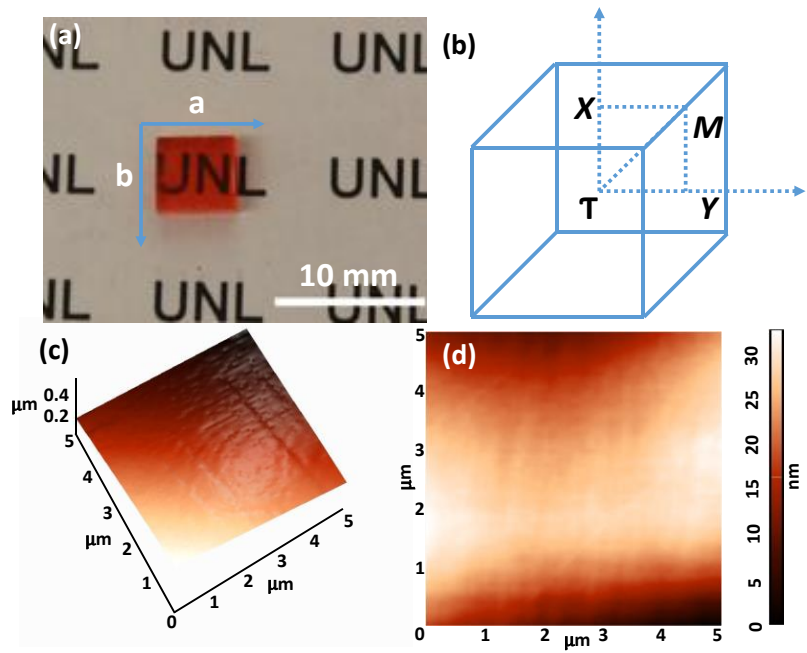

Fig. 1. (a) CH$_3$NH$_3$PbBr$_3$ single crystal with a surface area of 6 mm × 6 mm. (b) The reciprocal lattice of cubic CH$_3$NH$_3$PbBr$_3$. Three-dimensional (3D) (c) and 2D (d) AFM images of the crystal surface at two test areas (5 μm × 5 μm).

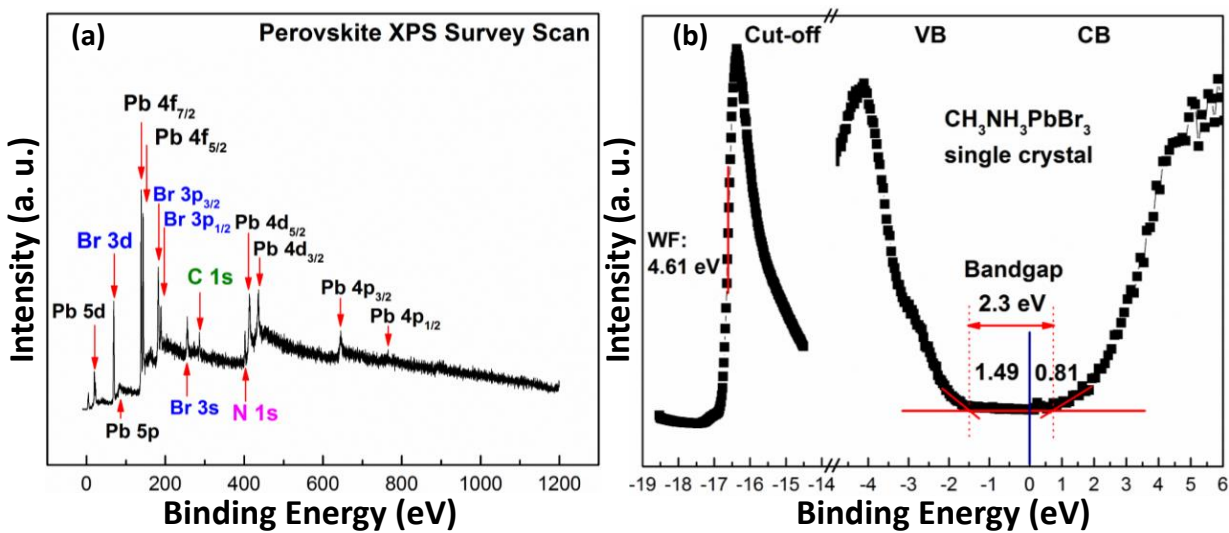

Fig. 2. (a) High resolution XPS survey scan and (b) Bandgap diagram of CH$_3$NH$_3$PbBr$_3$ single crystal.

Fig. 2a shows the high resolution XPS survey scan of the perovskite single crystal. The detailed photoelectron spectroscopy figures for individual elements could be found in Fig. S4. The composition of the as-cleaved sample is C: N: Pb: Br: O = 1.46: 1.05: 1.02: 3.04: 0.05, if taking Pb 4$f$ from perovskite as the basis (the extra 0.02 Pb 4$f$ is metal lead). This ratio is very close to the ideal value. The excessive carbon, nitrogen, bromine, and oxygen may have come from residual reactants used during the crystal growth process. The detailed analysis is shown in Table

S1. Another possible explanation is that the cleavages were done along an impurity rich region (where a fracture is easier to take place), but the bulk of the crystal still follows the stoichiometric ratio.

Fig. 2b is the bandgap diagram of the sample measured by UPS and IPES. It shows the binding energy (BE) from the Fermi level ($E_F$) of the material. For visual clarity, we normalized all the spectra to the same height. The cut-off energy is determined by the inflection point of the sharp change region of the cut-off spectrum.[29-31] The vacuum level (VL) is obtained from the difference between the photon energy (21.22 eV) and the cut-off energy. The VL of $CH_3NH_3PbBr_3$ single crystal is measured to be 4.61 eV above the $E_F$, i.e., the work function (WF). In the highest VB lying regions, the valence band maximum (VBM) of the single crystal displays ~1.49 eV. To obtain the detailed information on the unoccupied states of perovskite single crystal, we further collected the IPES data. The conduction band minimum (CBM) is measured to be ~0.81 eV above the $E_F$, while the VBM is ~1.49 eV below the $E_F$, corresponding to a bandgap of 2.3 eV that is consistent with previous reports[32, 33]. The valence band (VB) and conduction band (CB) edges are obtained using linear extrapolation as illustrated in our previous study[34]. The ionization potential (energy of the VBM referenced to the VL) is 6.10 eV. The crystal surface presents a slight n-type semiconductor behavior. The indirect character of the bandgap as discussed for $CH_3NH_3PbI_3$ thin films[35] was not observed for $CH_3NH_3PbBr_3$ single crystal.

In Fig. 3a and 3b, we present angle-resolved energy distribution curves (EDCs) of overall valence band as a function of electron emission angle $\theta$ relative to the surface normal measured along $\Gamma X$ and $\Gamma M$ directions, respectively. All the spectra were normalized to the same height and fitted by four Gaussian peaks, VB-A, VB-B, VB-C, and VB-D. One detailed analysis is shown in the inset of Fig. 3a. Both series of spectra show a structured and angle-dependent valence-band emission. For both of the $\Gamma X$ and $\Gamma M$ directions, as $\theta$ increases, all the peaks shift towards a higher BE until they reached the $\Gamma$ point (0 °). After passing through the $\Gamma$ point, the peaks show an upward dispersion and reach the minima at around 16.5 ° and 22.5 ° in $\Gamma X$ and $\Gamma M$ directions, respectively. This minima position corresponds to the boundary of the first Brillouin zone (BZ) in each direction. However, the shifts are not identical, which is somewhat stronger along $\Gamma M$ direction, being 0.41 eV, 0.18 eV, 0.20 eV, and 0.73 eV for VB-A, VB-B, VB-C, and VB-D in $\Gamma X$ direction; and 0.48 eV, 0.18 eV, 0.40 eV and 0.98 eV in $\Gamma M$ direction, separately. As $\theta$ increases beyond the 1st BZ boundary, all the peaks start to shift back toward higher BE again. These results indicate the nice symmetry around the BZ boundaries, as well as the $\Gamma$ point.

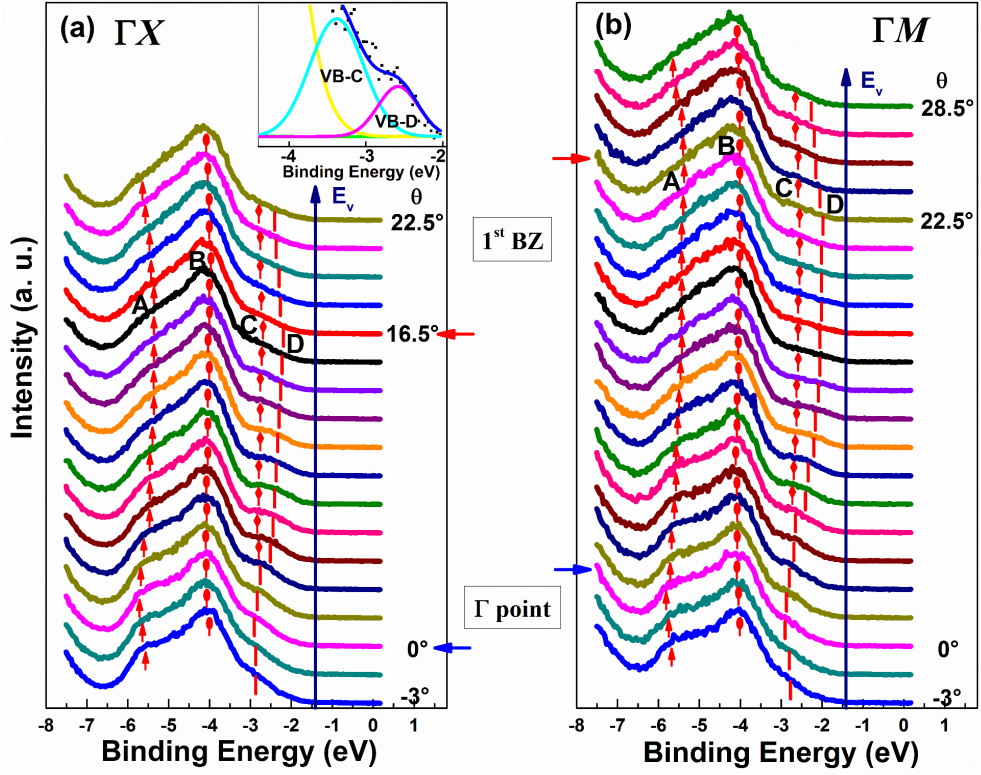

Fig. 3. (a) Angle-resolved energy distributions curves measured along ΓX (a); (b) Same, but measured along ΓM. The four VB peaks are denoted by arrows and bars. The inset shows the detailed fitting of VB-C and VB-D peaks.

The photon energy $h\upsilon$ was fixed at 21.22 eV. By adjusting the electron emission angle ($\theta$), the band dispersions $E(\mathbf{k})$ could be obtained simply from the measured values of kinetic energy $E_{kin}$ and the equation as follows[36]:

$$E_{kin} = h\upsilon - \phi - |E_B| \qquad (1)$$

$$\hbar k_\| = \sqrt{2mE_{kin}} \cdot \sin\theta \qquad (2)$$

$$\hbar k_\| = [2m(h\upsilon - \Phi - |E_B|)]^{\frac{1}{2}} \sin\theta \qquad (3)$$

Here $\hbar k_\|$ is the component parallel to the surface of electron crystal momentum, $\Phi$ = 4.61 eV is the work function, $E_B$ is the binding energy.

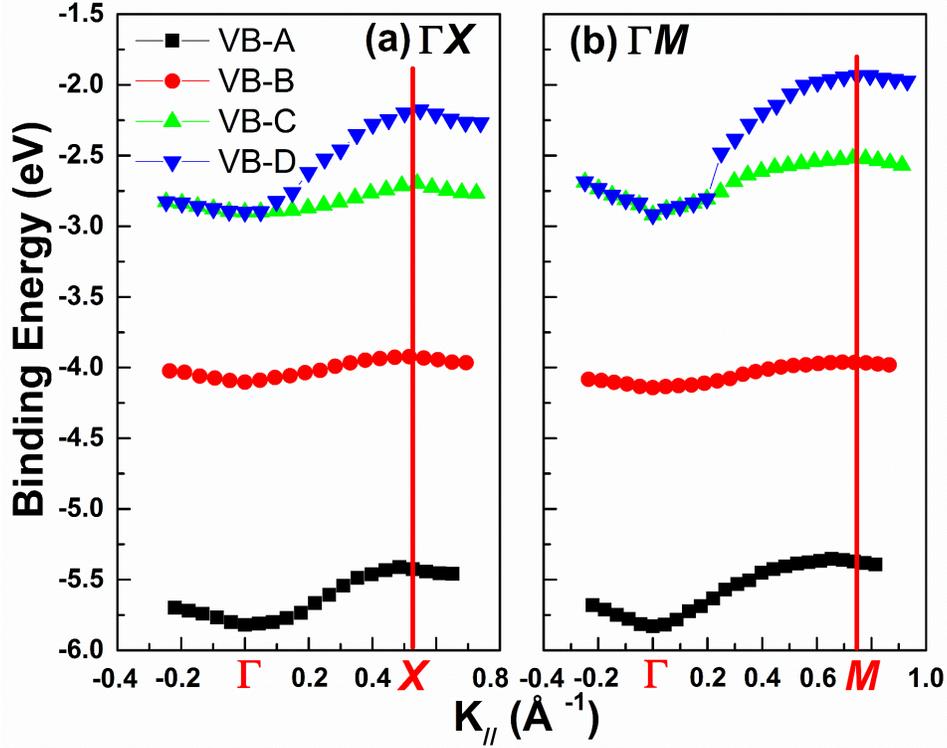

Fig. 4. Dispersions of the four valence bands of $CH_3NH_3PbBr_3$ single crystal along (a) ΓX and (b) ΓM directions. The red lines denote the first BZ boundaries.

Fig. 4 shows the energy dispersions of the four valence bands as a function of $k_\parallel$ along ΓX and ΓM directions. By comparison, the 1$^{st}$ BZ boundary has a peak energy of ~-1.94 eV at the M point that is higher than that of ~-2.18 eV at the X point. The experimental results are consistent with the first-principles calculations, being ~-0.90 eV and ~-1.70 eV at the M point and the X point, respectively.[21] It suggests that the VB is a mixture of Pb-s and Br-p orbitals[21,26], and the Br-p orbitals contribute mostly to the band features.[21,26] Interestingly, the VBM split into two bands, VB-C and VB-D, at about $k_\parallel$ =0.14 Å$^{-1}$ and ~0.24 Å$^{-1}$ for ΓX and ΓM directions, respectively. This band splitting could be explained by comparing with the DFT results from Jishi et al[21]. From the calculation, the bandgap of the cubic $CH_3NH_3PbBr_3$ single crystal is 2.23 eV and occurs at point R (1/2, 1/2, 1/2) in the BZ. The VBM dispersion in ΓX direction is ~0.56 eV, which is smaller than the measured one of 0.73 eV as shown in Fig. 4 (a). The VBM dispersion in ΓM direction is ~1.45 eV, which is larger than the measured value of 0.98 eV as shown in Fig. 4(b). According to the band structure calculation, the second and third valence bands are very close to each other, and almost have the same value with the VBM at the vicinity of the Γ point in both directions. The average dispersion of the two bands is ~0.14 eV and ~0.2 eV, which are comparable with the 0.20 eV and 0.40 eV of VB-C in ΓX and ΓM directions, respectively. Therefore, these two bands may contribute to the VB-C, and share the same $k_\parallel$ value with VB-D near the Γ point.

Noticeably, the $k_\parallel$ value at the X point is ~0.53 Å$^{-1}$ given by

$$k_\parallel = \frac{\pi}{a} = 0.53 \text{ Å}^{-1} \quad (4)$$

for the lattice parameter $a$ of 5.93 Å[37, 38]. The $k_\parallel$ value at the M point is ~0.75 Å$^{-1}$.

The measured band dispersions are the key factors to understand the transport characteristics of perovskite single crystals. Under the one-dimensional tight-binding approximation, the energy dispersions were fitted by

$$E_B = E_C - 2t \cos(ak_\parallel) \quad (5)$$

where $E_C$ is the binding energy of the band center, $t$ is the transfer integral, and $a$ is the lattice spacing. From the formulation, $t$ is estimated as 0.18 eV for the VB-D curve in ΓX direction. The effective hole mass $m_h^*$ at the VB region is given by

$$m_h^* = \frac{\hbar^2}{2ta^2} \quad (6)$$

Thus, the effective mass of VB-D band in ΓX direction is derived to be 0.59 $m_0$ ($m_0$ is the free electron mass). In a broad band model, where the band width W is much larger than $k_B T$ (26 meV), the lower limit drift mobility of the hole is estimated to be 33.90 cm$^2$V$^{-1}$s$^{-1}$ according to the equation[39]:

$$\mu_h > 20 \frac{m_0}{m_h^*} \times \frac{300}{T} \quad (7)$$

The calculated value of the hole mobility is comparable with some reports as ranging from 19.4 to 56.1 cm$^2$V$^{-1}$s$^{-1}$ by photoluminescence measurements[40], and 24.0 cm$^2$V$^{-1}$s$^{-1}$ from dark current-voltage measurements[41]. Other results such as 217 cm$^2$V$^{-1}$s$^{-1}$ and 206 cm$^2$V$^{-1}$s$^{-1}$ from time-of-flight measurements[5] are larger than the value we measured. Shi *et al*[32] measured the mobility of the same sample with two methods, being 20 to 60 cm$^2$V$^{-1}$s$^{-1}$ by Hall effect measurements and 115 cm$^2$V$^{-1}$s$^{-1}$ by time-of-flight (TOF) measurements. There is a distribution of mobility, while TOF measurement gives the mobility of carriers with the highest mobility. However, it is also possible that the structure of perovskite single crystal may be too complicated to be fully presented by the tight-binding approximation fitting, thus underestimating the hole mobility.

Conclusion

In summary, we have studied the electronic structure of cubic perovskite single crystal (001) surface with ARPES and IPES. The high quality of the smooth surface was confirmed by AFM and SEM. The elemental composition was investigated by XPS with a ratio close to the ideal value. Highly reproducible dispersive features of valence bands were observed with nice symmetry at the BZ center and boundaries. The four VB peaks are composed of Pb-*s* and Br-*p* orbitals as theoretical models predicted, and had different band dispersion widths. The largest dispersion came from the lowest binding energy band, being ~0.73 eV and ~0.98 eV for ΓX and ΓM directions, respectively. The measured band dispersions correspond to an effective hole mass as ~0.59 $m_0$ and a lower limit of the hole mobility of 33.90 cm$^2$V$^{-1}$s$^{-1}$ from the tight-binding fitting.

Acknowledgments

The authors would like to thank the financial supports from National Science Foundation (Grant Nos. CBET-1437656 and DMR-1303742). Huang acknowledges the financial support from National Science Foundation under the award of OIA-1538893. Meng was supported by China 1000-Young Talents Plan and the National Natural Science Foundation of China Grant (11574402, 51502351). Technical supports from the Nanocenter in University of Rochester are highly appreciated.

Valence Band Dispersion Measurements of Perovskite Single Crystal with Angle-resolved Photoemission Spectroscopy

Congcong Wang, Benjamin R. Ecker, Yongli Gao[*]

Department of Physics and Astronomy, University of Rochester, Rochester, NY 14627, USA

Haotong Wei, Jinsong Huang

Department of Mechanical and Materials Engineering, University of Nebraska-Lincoln, Lincoln, Nebraska 68588, USA.

Jian-Qiao Meng

School of Physics and Electronics, Central South University, Changsha, Hunan 410083, PR China.

* ygao@pas.rochester.edu

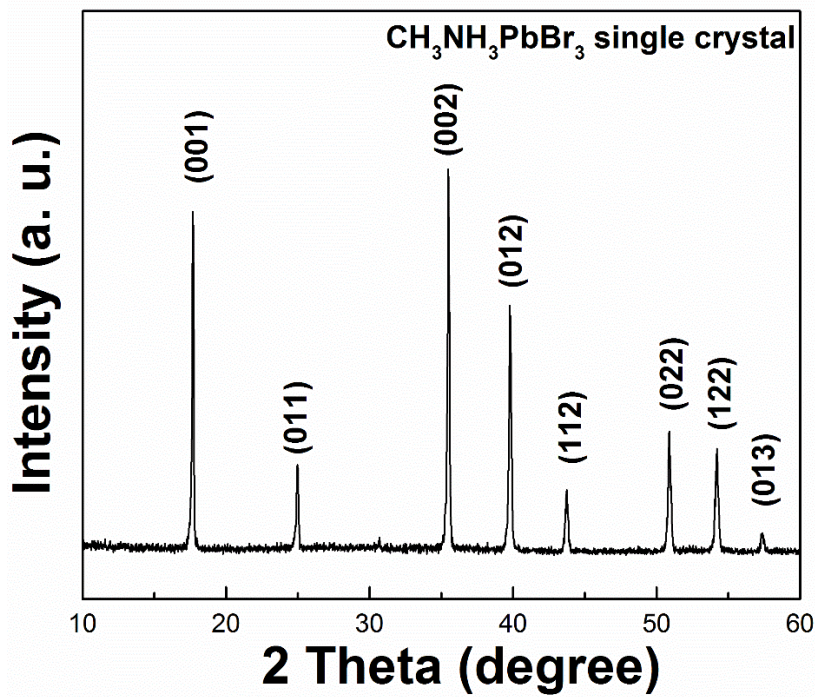

Figure S1. Powder x-ray diffraction result of $CH_3NH_3PbBr_3$ single crystal with multiple peaks corresponding to a clear cubic structure indexing.

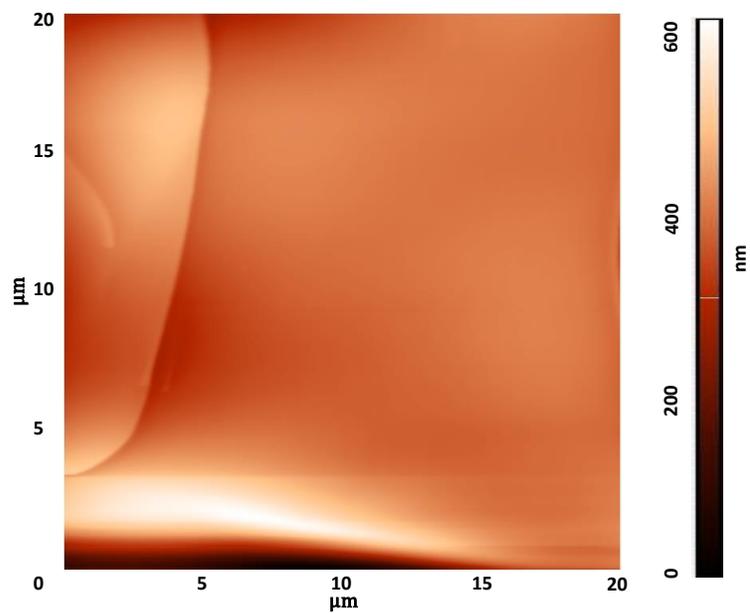

Figure S2. Two-dimensional atomic force microscopy (AFM) image of the crystal surface at the third position. The image size is 20 μm × 20 μm.

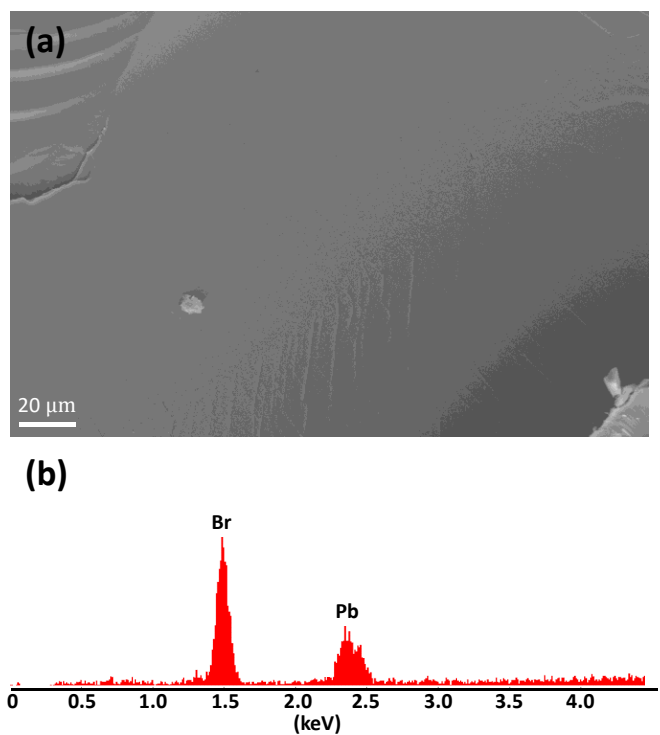

Figure S3. (a) Scanning electron microscope (SEM) image of the as-cleaved single crystal. (b) Energy dispersive spectroscopy (EDS) for the SEM imaged spot. The applied voltage is 20 KV. It clearly shows the signal of Pb and Br elements from the surface.

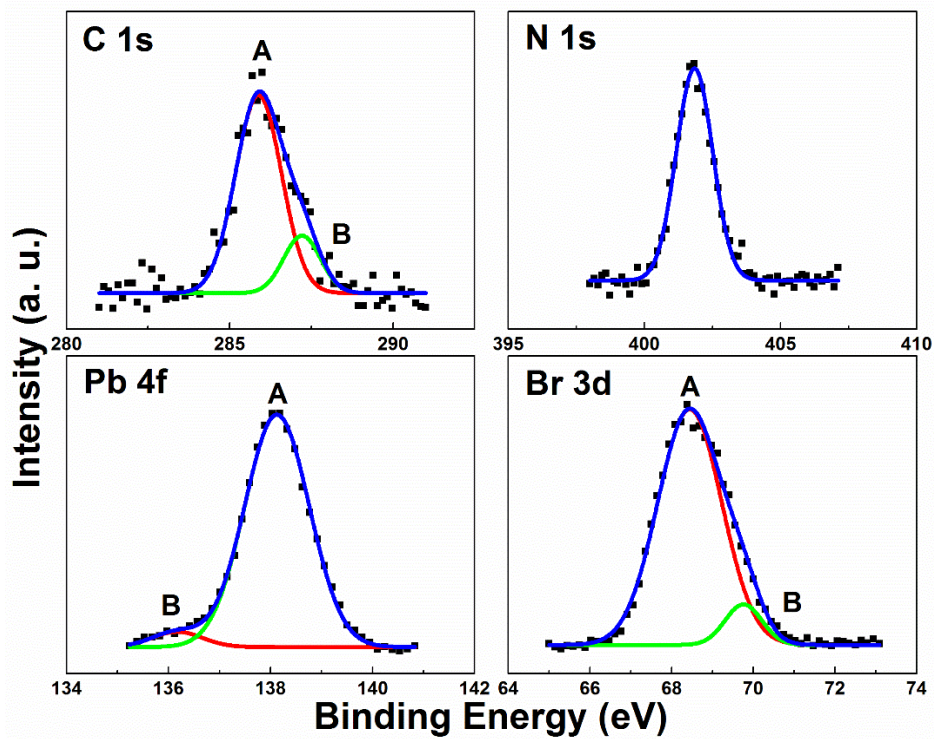

Figure S4. High resolution x-ray photoelectron spectroscopy (XPS) of the $CH_3NH_3PbBr_3$ single crystal. The C 1s-A peak is from perovskite, while the C 1s-B may be from the residual solvents, such as N, N-dimethylformamide (DMF) and dichloromethane (DCM), used during the crystal growth process. There is only one N 1s peak, indicating that the N 1s in DMF has the same binding energy as the N 1s in perovskite. The Pb 4f-A peak is from perovskite, while the Pb 4f-B peak is metal lead that may come from $PbBr_2$. The Br 3d-A is from perovskite, while Br 3d-B may come from $PbBr_2$. The O 1s peak is not shown here.

We obtained the areas of the XPS spectra of these elements by fitting Gaussian peaks after removing the secondary electron background, followed by normalization with corresponding atomic sensitivity factors. Therefore, the atomic ratio of perovskite single crystal is defined to be C: N: Pb: Br: O = 1.46: 1.05: 1.02: 3.04: 0.05, if taking the area of Pb 4f-A as the basis (the extra 0.02 is from Pb 4f-B). The ratio may suggest that the surface has $CH_3NH_3PbBr_3$: DMF: DCM: $PbBr_2$ = 1: 0.05: 0.31: 0.02 as listed in Table S1.

Table S1. Atomic Ratio of $CH_3NH_3PbBr_3$ Single Crystal.

| Element | C | N | Pb | Br | O |
|---|---|---|---|---|---|
| $CH_3NH_3PbBr_3$ | 1 | 1 | 1 | 3 | 0 |
| DMF | 0.15 | 0.05 | 0 | 0 | 0.05 |
| DCM | 0.31 | 0 | 0 | 0 | 0 |
| $PbBr_2$ | 0 | 0 | 0.02 | 0.04 | 0 |
| Total | 1.46 | 1.05 | 1.02 | 3.04 | 0.05 |